\documentclass[12pt,preprint]{aastex}

\usepackage[dvips]{color}

\newcommand{\dpp}[2]{\frac{\partial #1}{\partial #2}}

\newcommand{\mi}[1]{\mbox{\boldmath$#1$}}

\begin{document}
\title{Effect of suppressed excitation on the amplitude
distribution of 5-min oscillations in sunspots}

\author{Parchevsky, K.V., Kosovichev, A.G.}
\affil{W.W. Hansen Experimental Physics Laboratory, Stanford
University, Stanford, CA, USA}
\begin{abstract}
Five-minute oscillations on the Sun (acoustic and surface gravity
waves) are excited by subsurface turbulent convection. However, in
sunspots the excitation is suppressed because strong magnetic field
inhibits convection. We use 3D simulations to investigate how the
suppression of excitation sources affects the distribution of the
oscillation power in sunspot regions. The amplitude of random
acoustic sources was reduced in circular-shaped regions to simulate
the suppression in sunspots. The simulation results show that the
amplitude of the oscillations can be approximately 2-4 times lower
in the sunspot regions in comparison to the quiet Sun, just because
of the suppressed sources. Using SOHO/MDI data we measured the
amplitude ratio for the same frequency bands outside and inside
sunspots, and found that this ratio is approximately 3-4. Hence, the
absence of excitation sources inside sunspots makes a significant
contribution (about 50\% or higher) to the observed amplitude ratio
and must be taken into account in sunspot seismology.
\end{abstract}


\section{Introduction}
It has been known for a long time that 5-min solar oscillations have
significantly lower amplitude (by a factor of 2-5) in sunspots and
plages than in the quiet Sun \citep[e.g][]{Woods1981, Thomas1982,
Title1992}. \citet{Hindman1997} enumerated four possible mechanisms
to explain the observed power suppression: 1) reduction of
excitation of p-modes inside sunspots; 2) absorption of p-modes
inside sunspots \citep{Cally1995}; 3) the different height of
spectral line formation due to the Wilson depression; 4) altering of
p-mode eigenfunctions by the magnetic field. The precise
contribution of these effects to the observed amplitude reduction is
still unknown. In this paper we study the first effect: changes in
the oscillation amplitude due to suppression of acoustic sources by
using 3D numerical simulation of solar acoustic waves, which are
important for solar seismology studies. Inside sunspots, strong
magnetic field inhibits turbulent convective motions which are the
source of the 5-min solar oscillations. Therefore, the waves in the
5-min period (3 mHz frequency) range, observed in sunspots, mostly
come from the outside regions, and thus their amplitude is reduced
in comparison to the quiet Sun. Our goal was to estimate the
significance of this effect by modeling wave fields in horizontally
uniform background solar models with regions of reduced excitation.
The main result is that the suppression of oscillation sources
inside sunspots can make substantial (about 50\% or greater)
contribution to the reduction of amplitude inside sunspots, and thus
must be taken into account in sunspot seismology.

\section{Method}
Wave propagation on the Sun (in absence of magnetic field and flows)
can be described by the following system of linearized Euler
equations:
\begin{equation}\label{LinEuler}
\begin{array}{l}
\displaystyle\dpp{\rho'}{t} + \mi{\nabla}\cdot(\rho_0\mi{u}')= 0
\\[9pt]
\displaystyle\dpp{}{t}(\rho_0 \mi{u}') + \mi{\nabla}\;p' = \mi{g}_0
\rho' + \mi{f}(x,y,z,t),
\end{array}
\end{equation}
where $u',v',w'$ are the perturbations of $x,y,z$ velocity
components, $\rho'$ and  $p'$ are the density and pressure
perturbations correspondingly, $\mi{f}(x,y,z,t)$ is the function
describing the acoustic sources. The pressure $p_0$, density
$\rho_0$, and gravitational accelerations $\mi{g}_0$ of the
background reference model depend only on depth $z$. We used the
adiabatic relation $\delta\rho/\rho_0 = 1/\Gamma_1\; \delta p/p_0$
between Lagrangian variations of pressure $\delta p$ and density
$\delta\rho$. The adiabatic exponent $\Gamma_1$ was calculated from
OPAL equation of state \citep{Rogers1996}.

The numerical method is based on a high-order finite difference
scheme, developed by \citet{Tam1993} and described in details by
\citet{Parchevsky2006}. The coefficients of this finite difference
scheme are chosen to minimize the error of the Fourier transform of
numerical derivatives. Such a scheme preserves the dispersion
relations of the continuous case for shorter wavelengths. A
3rd-order strong stability preserving Runge-Kutta method
\citep{Shu2002} was used for time integration.

The standard solar model \citep{Christensen-Dalsgaard1996} with a
smoothly joined chromospheric model of \citet{Vernazza1976} was used
as the background state. The model was corrected in the near-surface
layers to prevent development of convective instability in the
superadiabatic layer by replacing large negative values of the
Brunt-V\"{a}is\"{a}l\"{a} frequency by zero (or small positive)
values and recalculating the hydrostatic equilibrium
\citep{Parchevsky2006}. This is a necessary modification for all
linear simulations of this type. It suppresses rapidly growing
unstable convective modes but does not affect essential properties
of acoustic wave propagation in the Sun.

To prevent reflection of acoustic waves from the boundaries of the
computational domain, we follow the PML (Perfectly Matching Layer)
method of \citet{Hu1996}. We set the top non-reflecting boundary
condition above the temperature minimum. This simulates a realistic
case when the waves are only partially reflected by the photosphere.
The waves with frequencies higher than acoustic cut-off frequency
$\nu_{ac}\sim 5$ mHz pass through the photosphere and are absorbed
by the PML. For frequencies below $\nu_{ac}$, the top boundary does
not affect the reflection because the acoustic waves are mostly
reflected by the photosphere and become evanescent in the
chromosphere.

The damping mechanism of solar modes below the acoustic cut-off
frequency is not yet completely understood. It can be due to wave
scattering on turbulence in subsurface layers \citep{Murawski2003}
and also due to partial escape of waves and radiative losses
\citep{Christensen-Dalsgaard1983}. We have investigated both of
these mechanisms. The atmospheric damping was modeled by imposing
the upper absorbing boundary at different levels, and choosing the
height of this boundary in such a way (about 500 km above the
photosphere) that the observed line widths in the oscillation power
spectrum are well reproduced. To model the subsurface (turbulent)
damping we followed \citet{Gizon2002} and added a friction-type term
$-\sigma(z)\rho_0 v_z$ to the vertical component of momentum
equation, where damping coefficient $\sigma(z)$ is constant above
the photosphere and smoothly decreases to zero at a depth of about
500 km. For this case, the upper boundary was placed at the
chromosphere-corona transition layer (about 1750 km above the
photosphere), and the value of $\sigma(z)$ was adjusted to match the
observed line widths and relative amplitude of the peaks in the
acoustic spectrum. The lateral boundary conditions are periodic.
Duration of simulations was 4.5 hours of solar time. We compared
results with longer runs and checked that the root mean square (rms)
oscillation amplitude reaches an equilibrium state, and also that
the simulated acoustic spectra are close to the observed power
spectrum.

\begin{figure}
\includegraphics[width=\textwidth]{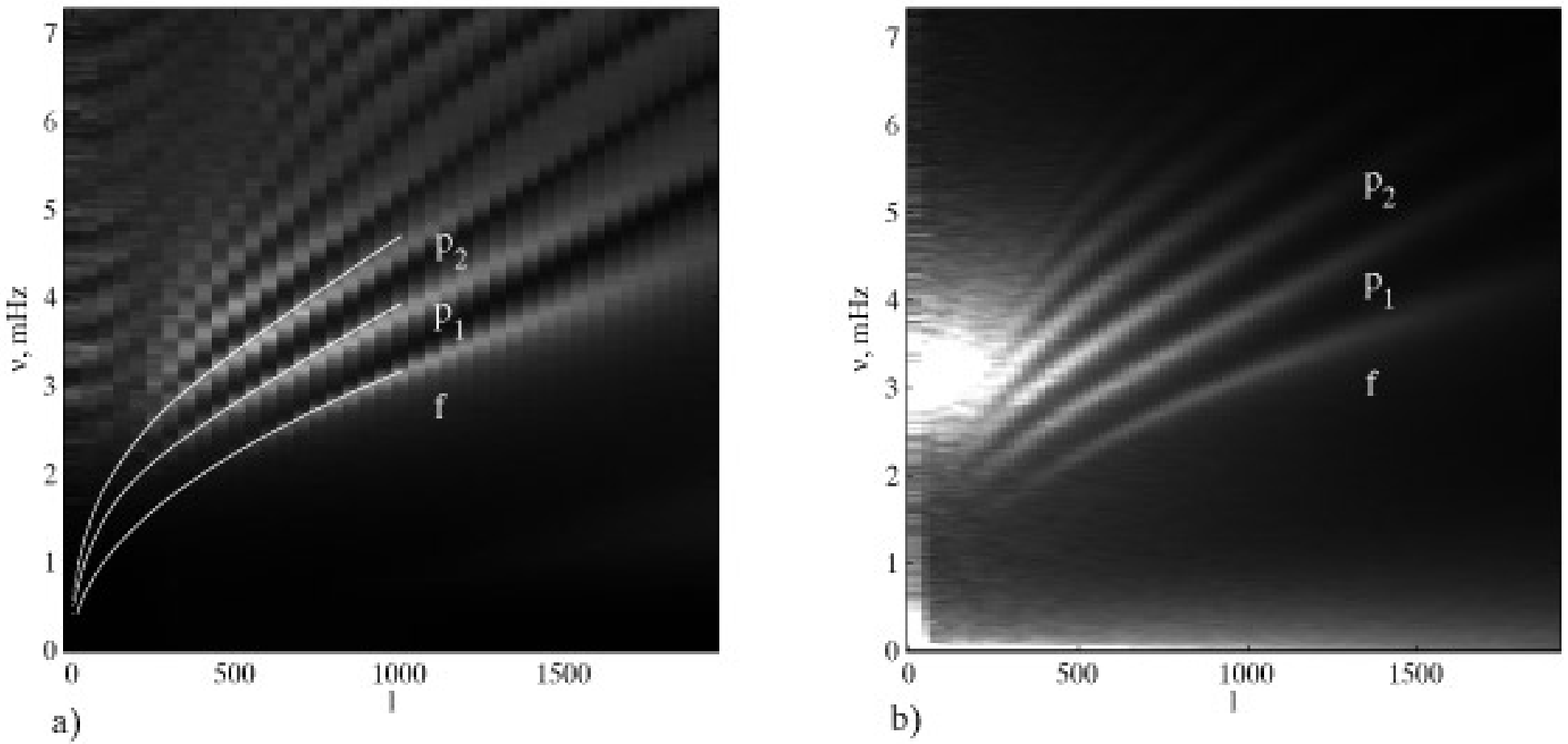}
\caption{Power spectrum of the vertical component of velocity
obtained from the simulations a) and high-resolution SOHO/MDI data
b) \citep{Scherrer1995}. The acoustic spectral density depicted by a
gray-scale is given in arbitrary units. The white curves show
observed $f$, $p_1$, and $p_2$ mode ridges; $l$ is the angular
degree.} \label{f1}
\end{figure}

The question about the depth of the acoustic sources on the Sun is
also still open. Numerical simulations of solar convection
\citep{Stein2001} show that in the region around $3\div4$ mHz the
most driving occurs between the photoshpere and sub-surface layer
500 km deep, with a maximum driving at 200-300 km below the surface.
Accordingly, the sources were randomly distributed in time and on a
horizontal plane 350 km below the photosphere. We considered also
the case of shallow (100 km deep) sources, as suggested by
\citet{Nigam1999}, \citet{Kumar2000}. The sources were modeled by
spherically symmetric Gaussian shape vertical force perturbations
with FWHM of 300 km and random amplitudes and frequencies. The time
dependence was either a one-period sin function:
$\sin[\omega(t-t_0)],\; t_0\leq t\leq t_0+2\pi/\omega$ or Ricker's
wavelet: $(1-2x^2)e^{-x^2},\;$ where $x= [\omega (t-t_0)/2-\pi],\;
t_0\leq~t\leq~t_0+4\pi/\omega$. The frequency distribution of
acoustic sources was uniform in the range of $2\div8$~mHz. The
simulations were carried out in the rectangular domain of size
122$\times$122$\times$30 $\mbox{Mm}^3$ using a uniform
816$\times$816$\times$630 grid. The simulated and observed power
spectra are shown in Fig.~\ref{f1}.

\section{Results}

\begin{figure}
\includegraphics[width=\textwidth]{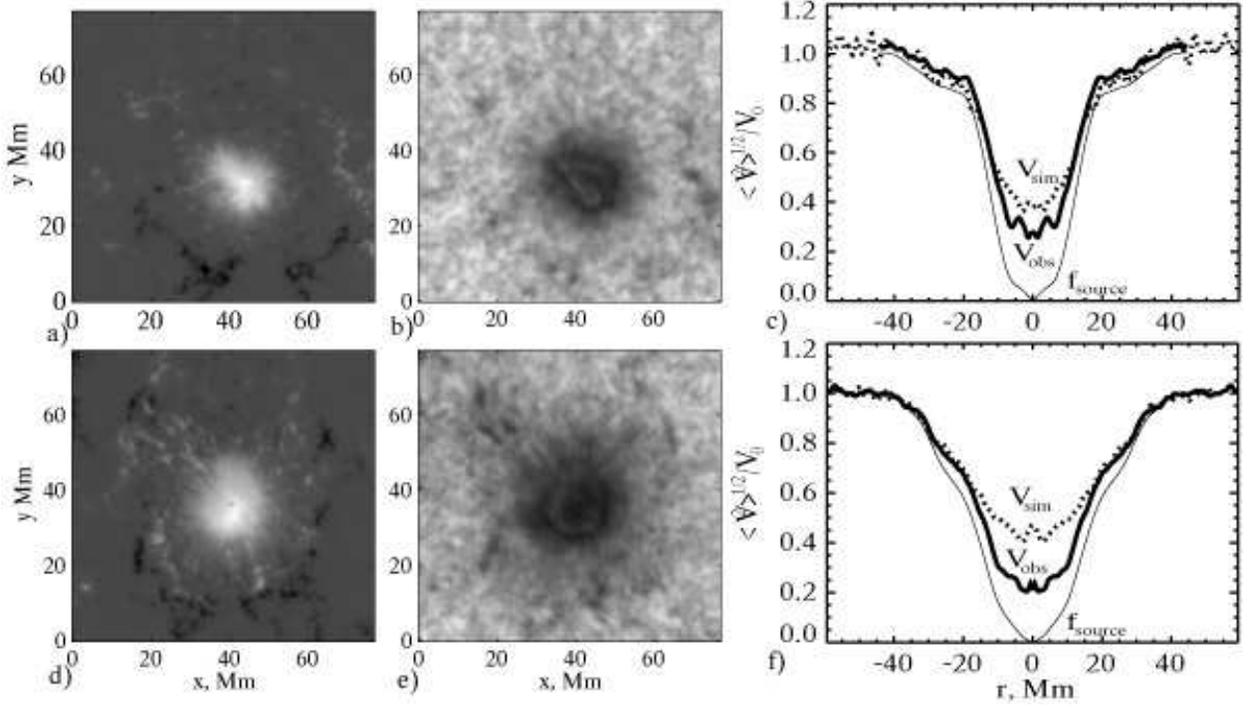}
\caption{Panels a) and d) represent the line-of-sight magnetic field
maps, panels b) and e) show the oscillation amplitude maps. The
profiles of rms oscillation velocities at frequency 3.65 mHz for
observations (thick solid curves), simulations (dashed curves), and
the profiles of source strength (thin solid curves) for sunspots in
two active regions AR10373 (top) and AR8243 (bottom) are shown in
panels c) and f).} \label{f2}
\end{figure}

Using this method we simulated the distribution of oscillation power
for sunspots of various size, acoustic source models, and compared
with observations. Observations for sunspots in active regions
AR10373 (a,b,c) and AR8243 (d,e,f) obtained by SOHO/MDI are shown in
Fig.~\ref{f2}. Panels a) and d) represent maps of the line-of-sight
magnetic field from high-resolution MDI data. Panels b) and e) show
corresponding vertical velocity oscillation amplitude maps, averaged
in the frequency interval $\Delta\nu=1.2$ mHz with central frequency
$\nu=3.65$ mHz. The azimuthally averaged profiles of the oscillation
amplitude (thick solid curves for observations and dashed curves for
simulations) and the profile of the source strength (thin solid
curves) are shown in panels c) and f).

\begin{figure}
\includegraphics[width=\textwidth]{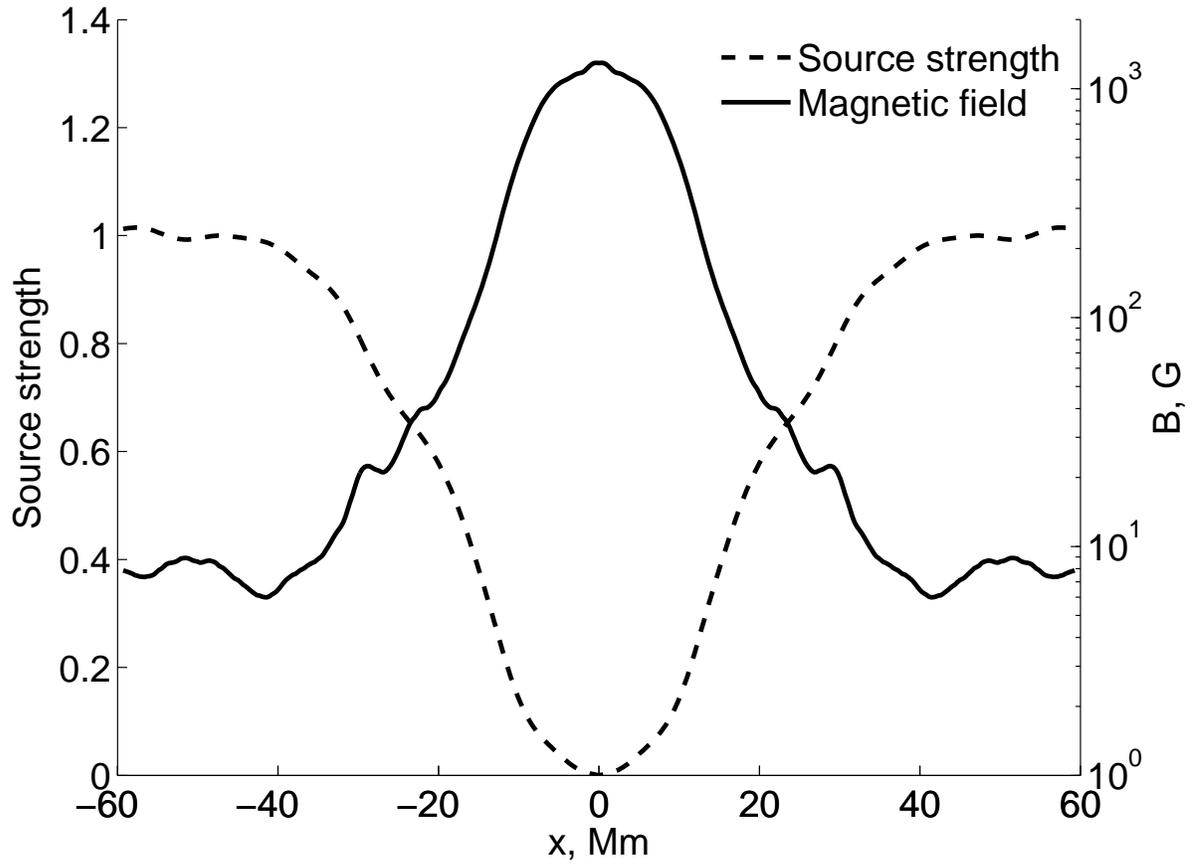}
\caption{Horizontal profiles of the source strength (dashed curve)
and angular averaged magnetic field strength (solid curve).}
\label{f3}
\end{figure}

We simulated the suppression of the acoustic sources (changes in
acoustic emissivity) due to magnetic field in sunspots by zeroing
the source amplitude at the center of sunspot umbra and smoothly
increasing the source strength in the penumbra towards a constant
value outside the sunspots (source masking). Numerical experiments
with different profiles of the source strength show in all cases
that the simulated wave field profile has the shape similar to the
profile of the acoustic source strength (this was not a priori
obvious). Thus, we use a horizontal profile of the observed wave
field (shifted and scaled to be in range [0,1]) as an acoustic
source strength profile. In reality the strength of acoustic sources
depends on magnetic field strength. The actual dependence is
unknown, however, the distribution of acoustic sources calculated
from the wave field  is similar to the inverse profile of the
sunspot magnetic field (Fig.~\ref{f3}).

The simulation results (Fig.~\ref{f2} c,f) show that the waves
propagate into the region of reduced excitation, but oscillation
amplitude is substantially decreased. However it still does not
match the observed values at sunspot center. This means that the
suppression of acoustic sources is obviously a very important effect
in sunspot seismology. However, other factors are likely to play
role too.

\begin{figure}
\includegraphics[width=\textwidth]{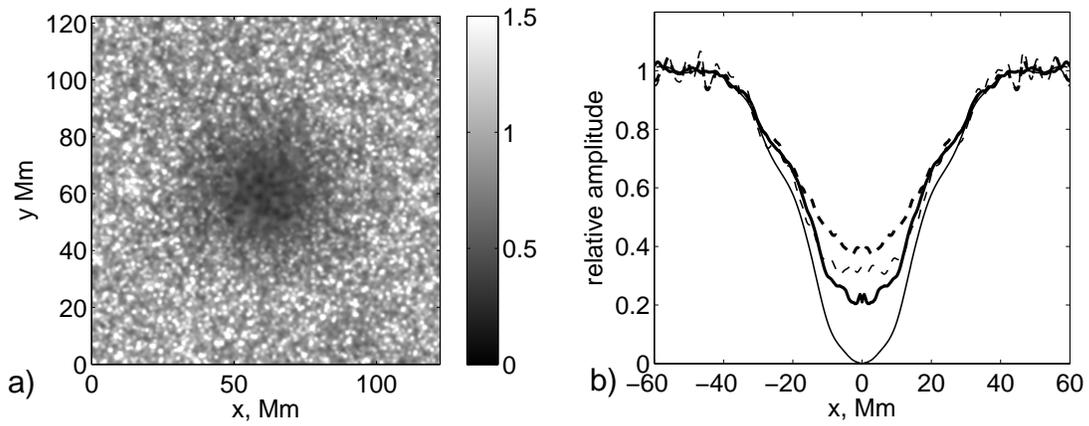}
\caption{Vertical velocity oscillation amplitude map for the
atmospheric damping a).  Angular averaged amplitude profiles
obtained from simulations are shown in panel b) by the thick dashed
curve for atmospheric damping and the thin dashed curve for explicit
damping. The thin solid curve represents the source strength profile
calculated from observations of the sunspot in AR8243. The thick
solid curve shows the observational amplitude profile for the same
sunspot.} \label{f4}
\end{figure}

The result of simulations for the sunspot in active region AR8243 is
shown in Fig.~\ref{f4}. Panel a) represents the amplitude map for
simulations with a low (500 km) top absorbing boundary without
explicit damping and filtered with a spatial Gaussian filter with
FWHM of 1.5 arcsec to reflect the instrumental smoothing. Panel b)
represents the angular averaged amplitude profiles for observations
(thick solid curve); simulations with a low top absorbing boundary
without explicit damping (thick dashed curve); simulations with a
high (1750 km) top absorbing boundary with an additional damping
term in the subsurface layers, introduced into the z-component of
the momentum equation (thin dashed curve). The thin solid curve
shows the strength profile of the acoustic sources. It was shown by
\citet{Jones1989} that the hight of formation of Ni I 6768 \r{A}
line used by the MDI instrument is in the range of 200$\div$300 km
above the photosphere. To match the simulations with the MDI
observations, the amplitude maps were plotted for the height of 300
km above the photosphere. Velocities $V_{in}$ and $V_{out}$ outside
and inside the masked region were calculated by spatial averaging of
the amplitude map in the circle of radius 9.0 Mm around the center
of the sunspot and outside the sunspot in the ring with inner and
outer radii of 45 Mm and 50 Mm correspondingly. The ratios
$V_{out}/V_{in}$ are $3.2\pm 1.1$ for simulations with explicit
damping and $2.3\pm 0.4$ for simulations without explicit damping.
The amplitude ratio obtained from the observations (for the same
averaging regions) equals $3.9\pm 1.0$. Thus, in both cases our
simulations show that more than a half of the amplitude suppression
comes from the absence of acoustic sources inside the sunspot, and
the results are not much sensitive to the mechanism of wave damping.
Our numerical experiments showed that amplitude ratio
$V_{out}/V_{in}$ is insensitive to the detailed shape of the profile
of the acoustic source strength. However, it is important that the
FWHM of the source strength is close to the FWHM of the averaged
horizontal amplitude profile of the observed wave field. We found
that the amplitude ratio increases for smaller depth $h_{src}$ of
the acoustic sources and equals $5.5\pm 1.7$ for $h_{src} = 100$ km
below the photosphere (low top absorbing boundary without explicit
damping). This can be an evidence that the depth of the acoustic
sources is between 100 km and 350 km, if the absence of acoustic
sources is a dominating mechanism of the amplitude suppression in
sunspots.

\begin{figure}
\begin{center}
\includegraphics[width=\textwidth]{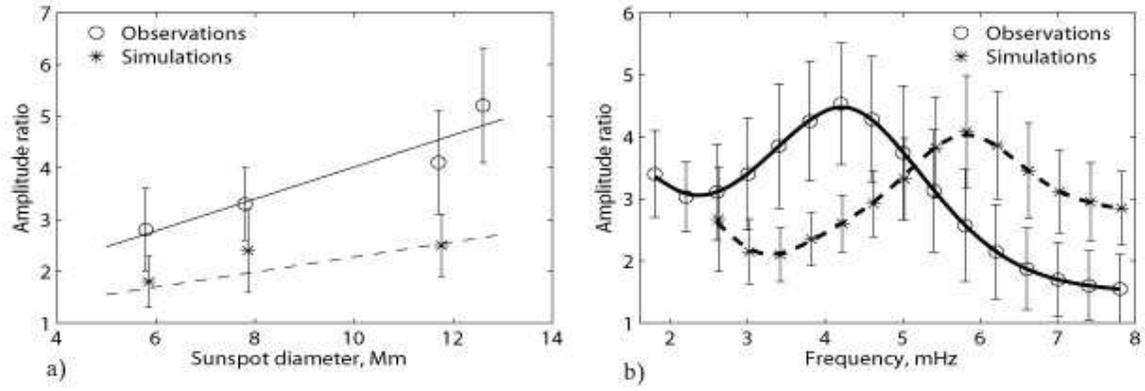}
\end{center}
\caption{Ratio of the oscillation amplitudes of the vertical
velocity $V_{out}/V_{in}$ as a function of umbra diameter a) and
frequency dependence of the amplitude ratio for the sunspot in
AR8243 b). } \label{f5}
\end{figure}

We have carried such simulations for several other sunspots, and
obtained similar results. The comparison between simulations and
observations for sunspots of different sizes is shown in
Fig.~\ref{f5}a. The ratio $V_{out}/V_{in}$ as a function of umbra
diameter is plotted. The open circles represent observations, the
stars represent simulations. Both the simulations and the
observations show the same trend: the amplitude suppression
increases with the size of sunspots. On average, the amplitude ratio
is about a half of the observed one for the sources at depth 350 km,
and higher for shallower sources. Frequency dependence of the
amplitude ratio for AR8243 is shown in Fig.~\ref{f5}b. The frequency
shift between the simulations and observations in Fig.~\ref{f5}b can
be an evidence that the acoustic cut-off frequency inside sunspots
is much lower that in the quiet Sun, but this requires further
investigation. This may be one of the additional factors which
affect the oscillation amplitude in sunspots.

\section{Discussion}
We have carried numerical simulations of the effect of reduced
excitation of solar oscillation in sunspot regions. The oscillations
are excited by random sources, modeled as vertical momentum and
pressure perturbations (in reality caused by turbulent convection).
In sunspot regions, the wave sources are weaker because the magnetic
field of sunspots inhibits convective motions. The results of
simulations show that for a wide range of sunspot diameters more
than a half of suppression of oscillation amplitude can be explained
by the absence of acoustic sources in sunspots.

Our simulations also showed that the oscillation amplitude in
regions of suppressed excitation only weakly depends on the wave
damping mechanism in the upper convection zone and atmosphere as
long as the line widths in the simulated power spectrum are close to
the observed ones. We modeled wave damping by two methods:
introducing a friction-type term into the z-component of the
momentum equation and putting the wave absorbing boundary at various
heights in the chromosphere. In both cases we get the similar ratios
of oscillation amplitudes. If the acoustic cut-off frequency in
sunspots is reduced this will increase the damping and may lead to
even stronger reduction of the wave amplitude compared to the quiet
Sun.

We thank Prof. P. Scherrer for fruitful discussions.

\clearpage

\clearpage

\clearpage

\clearpage

\clearpage

\end{document}